\documentclass[11pt]{article} 
\usepackage{hyperref}
 \pdfoutput=1
 \usepackage{hyperref} %
 \pdfoutput=1
\begin{document}
 \title{Surface explosion cavities}
 
\author{Adrien Benusiglio$^{1,2}$, David Qu\'er\'e$^2$, Christophe Clanet$^1$ \\
  \\\vspace{6pt}
  {\small \emph{$^1$Ladhyx, \'Ecole Polytechnique, France} }\\
   {\small \emph{$^2$Physique et M\'ecanique des Milieux H\'et\'erog\`enes, ESPCI, France}} }

\maketitle
\begin{abstract} 
 We present a \href{run:/anc/Surface_Explosion_Cavities_Large.mpg}{fluid dynamics video} on cavities created by explosions of firecrackers at the water free surface. We use three types of firecrackers containing 1, 1.3 and 5 g of flash powder. The firecrackers are held with their center at the surface of water in a cubic meter pool. The movies are recorded from the side with a high-speed video camera.

 Without confinement the explosion produces an hemispherical cavity. Right after the explosion this cavity grows isotropically, the bottom then stops while the sides continue to expand. In the next phase the bottom of the cavity accelerates backwards to the surface. During this phase the convergence of the flow creates a central jet that rises above the free surface. 
 
In the last part of the video the explosion is confined in a vertical open tube made of glass and of centimetric diameter. The explosion creates a cylindrical cavity that develops towards the free end of the tube. Depending on the charge, the cavity can either stop inside the tube or at its exit, but never escapes.

\end{abstract}

 \end{document}